\documentclass[twocolumn,showpacs,preprintnumbers,amsmath,amssymb]{revtex4}   
\usepackage{epsfig}
\usepackage{dcolumn}   
\usepackage{bm}        
\usepackage{amssymb}
\newcommand{\bald}[1]{{\bf #1}}

\begin{document}

\title{Anomalous Conical Di-jet Correlations in pQCD vs AdS/CFT}
\author{Barbara Betz$^{1,2}$, Miklos Gyulassy$^{1,3,4}$, Jorge Noronha$^3$, and Giorgio Torrieri$^{1,4}$}
\affiliation{$^1$Institut f\"ur Theoretische Physik, Goethe Universit\"at, Frankfurt, Germany\\
$^2$Helmholtz Research School,  Goethe Universit\"at, GSI and FIAS, Germany\\
$^3$Department of Physics, Columbia University, 538 West 120$^{\,th}$ Street, New York,
NY 10027, USA\\
$^4$Frankfurt Institute for Advanced Studies, Frankfurt, Germany}

\begin{abstract}
We propose an identified heavy quark jet observable
to discriminate between weakly coupled pQCD and
strongly coupled AdS/CFT models of quark gluon plasma dynamics in ultra-relativistic nuclear collisions at RHIC and LHC energies.
These models are shown to predict qualitatively different associated hadron
correlations with respect to tagged heavy quark jets.
While both models feature  similar far zone Mach and  diffusion wakes,
the far zone stress features are shown to be too weak to survive
thermal broadening at hadron freeze-out.
However, these models differ significantly in a near zone ``Neck'' region
where strong chromo-fields
sourced by the heavy quark jet couple to the polarizable
plasma. Conical associated
correlations, if any, are shown to be dominated by
the jet induced transverse flow in the
Neck zone and unrelated to the weak far zone wakes.
Unlike in AdS/CFT, we show that the induced transverse flow in the Neck zone is too weak in pQCD to produce conical correlations
after Cooper-Frye freeze-out. The observation of conical
correlations violating Mach's law would favor the strongly-coupled
AdS/CFT string drag dynamics, while their absence would favor
weakly-coupled pQCD-based chromo-hydrodynamics.
\end{abstract}

\date{\today}
\maketitle

\section{Introduction}\label{intro}

Recent interest in Mach-like conical di-jet correlations
\cite{Adler:2005ee} is due to suggestions
\cite{Stoecker:2004qu,shuryakcone} that a measurement of the
dependence of the cone angle on the supersonic jet velocity $v$ could
provide via Mach's law ($\cos\phi_M= c_s/v$) a constraint on the
average speed of sound in the strongly coupled Quark-Gluon Plasma
(sQGP) \cite{Gyulassy:2004zy} created at the Relativistic Heavy Ion
Collider (RHIC). In Refs.\ \cite{NGTnonmach} we explored the
robustness of this interpretation using the string drag model of
strongly coupled plasma-field interactions
\cite{Herzog:2006gh,gubsermach} in the context of the Anti-de
Sitter/Conformal Field Theory (AdS/CFT) correspondence
\cite{maldacena}. This AdS/CFT motivated model provides a detailed
holographic description of the induced stress tensor
\cite{Friess:2006fk} in the wake of a heavy quark jet moving at a
constant velocity through a static strongly-coupled $\mathcal{N}=4$
Supersymmetric Yang-Mills (SYM) background plasma at finite
temperature $T_0$. Direct tests of this AdS/CFT string drag model
using the ratio of bottom to charm nuclear modification factors in
high energy nuclear collisions at RHIC and LHC have been proposed in
Ref. \cite{Horowitz:2007su}. In this work, we concentrate on another
observable: the hadron correlations associated with tagged identified
heavy quark jets.

The AdS/CFT stress solution for a supersonic heavy quark
\cite{gubsermach,Chesler:2007an} features the expected far zone Mach
cone as well as a strong forward moving diffusion wake.  The far zone
response is well described in the strong coupling limit of an
$\mathcal{N}=4$ SYM plasma by a ``minimal'' shear viscosity over
entropy density ratio $\eta/s=1/4\pi$ \cite{Policastro:2001yc} (near
the uncertainty principle limit \cite{Danielewicz:1984ww}). However,
in \cite{gubsermach,Gubser:2007ni} it was noted that the strong
forward diffusion wake in the far zone of the AdS/CFT solution could
spoil the double-shoulder signature of the Mach wake in accord with
the general discussion of far zone hydrodynamics in
\cite{shuryakcone}.

In \cite{NGTnonmach} it was shown that in the strict supergravity
limit, $N_c\gg 1,\, g_{SYM}^2\ll 1$ but $\lambda=g_{SYM}^2 N_c\gg 1$,
in fact the far zone wakes have such small amplitudes that they only
lead to a single broad peak in the away-side hadronic correlation
after Cooper-Frye (CF) freeze-out of the fluid \cite{Cooper:1974mv}.
However, the AdS/CFT string drag solution features a novel
nonequilibrium ``Neck'' near zone, where especially strong transverse
flow relative to the jet axis induces an apparent conical
azimuthal correlation of associated hadrons even after CF thermal broadening at freeze-out.

In Ref.\ \cite{NGTnonmach}, the Neck zone was defined as the region
near the heavy quark where the local Knudsen number
\cite{Noronha:2007xe} is $K_N(X)= \Gamma_s \,{|\nabla\cdot {\bald
M}|}/{|{\bald M}|} >1/3$.  where $M^i (X)=T^{0i}(X)$ is the momentum
flow field of matter and $\Gamma_s\equiv 4\eta/\left(3sT_0\right)\ge
1/\left(3\pi T_0\right)$ is the sound attenuation length, which is
bounded from below for ultra-relativistic systems
\cite{Policastro:2001yc,Danielewicz:1984ww}.  In the Neck region the
induced transverse flow is surpringly large and is unrelated to the
far zone Mach's law. In AdS/CFT, this is the field-plasma coupling
zone where the stress tensor has a characteristic interference form
dependence on the coordinates, $O(\surd\lambda T_0^2/R^2)$
\cite{Yarom:2007ni,Gubser:2007nd}, with $R$ denoting the distance to
the heavy quark in its rest frame. In contrast, the stress in the far
zone has the characteristic $O(T_0^4)$ form.  In addition, very near
the quark the self Coulomb field of the heavy quark contributes with a
singular stress $O(\sqrt{\lambda}/R^4)$ \cite{Friess:2006fk}.

The above strong coupling AdS/CFT results motivated us in the present
work to study whether similar novel near zone field-plasma dynamical
coupling effects arise in weakly coupled perturbative Quantum
Chromodynamics (pQCD). In Refs.\ \cite{Neufeld:2008fi,
Neufeld:2008hs,Neufeld:2008dx} the heavy quark jet induced stress in a
weakly-coupled QGP (wQGP in contrast to sQGP)
was computed analytically in the linear response approximation
based on the Asakawa-Bass-M\"uller (ABM) \cite{Asakawa:2006tc,Asakawa:2006jn}
generalization of chromo-viscous hydrodynamics \cite{Heinz:1985qe}.
The ABM generalization concentrates
on the ``anomalous diffusion'' limit,
where the conductivity is dominated by field rather
than stochastic dissipative scattering dynamics.

As in the AdS/CFT string drag model, the generic
far zone  Mach and diffusion wakes are also clearly
predicted in the pQCD based ABM formulation
\cite{Neufeld:2008fi,Neufeld:2008dx} as we show for the case of a $v=0.9$
heavy quark jet in Fig.1.
However, the question of
whether the far zone Mach cone flow correlations
survives CF freeze-out of the plasma was not addressed up to now. In this
paper, we extend the work of \cite{Neufeld:2008fi,Neufeld:2008dx}
 by solving numerically the full nonlinear
3+1D relativistic hydrodynamic equations using the SHASTA hydro code \cite{Rischke:1995ir}, supplemented with the chromo-viscous stress
 source derived in Refs.\ \cite{Neufeld:2008fi,Neufeld:2008hs}.
We specialize to the
ideal fluid case of vanishing viscosity to minimize the dissipative
broadening of any conical correlations and therefore maximizing
the signal to noise ratio.

\begin{center}
\begin{figure}[th!]
\centerline
\centering
\epsfig{file=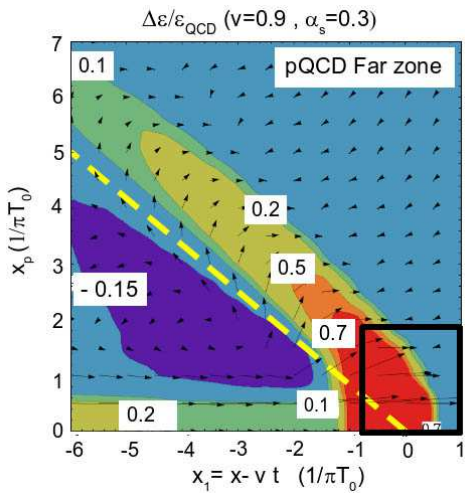,width=2.6in,clip=}
\caption{
(Color online)
The fractional energy density perturbation $\Delta \varepsilon/\varepsilon_0\equiv \varepsilon(x_1,x_p)/\varepsilon_0-1$ (in the lab frame)
due to a heavy quark with $v=0.9$ in a QCD plasma of temperature
$T_0=200$ MeV. The induced fluid stress was calculated using 3+1D hydrodynamics {\protect\cite{Rischke:1995ir}}
with the anomalous pQCD source of Neufeld \cite{Neufeld:2008hs} (left panel) and AdS/CFT \cite{NGTnonmach} (right panel). A trigger jet (not shown) moves in the $-\hat{x}$ direction. The away-side jet moves in $\hat{x}$ direction and contours of $\Delta \varepsilon/\varepsilon_{0}=-0.15,0.1,0.2,0.5,0.7$ are labeled in a comoving coordinate system with $x_1=x-vt$ and the transverse radial coordinate $x_p$ in units of $1/\pi T_0\approx 0.3$ fm after a total transit time $t=5$fm/c$=14.4/(\pi T_0)$. The ideal Mach cone for a point source is indicated by the yellow dashed line in the $x_1-x_p$ plane. See Fig. \ref{Stress} for a zoom of the Neck region inside of the black box.
}
\label{convention}
\end{figure}
\end{center}

We emphasize that our aim here is not to address the current light quark/gluon jet RHIC correlation data.
Our goal is to point out the significant differences between weakly coupled and strongly coupled models mechanisms of heavy quark energy loss
that can be tested experimental when identified heavy quark (especially bottom quark) jet correlations will become feasible to measure.
We limit this study to the most favorable idealized conditions (uniform static plasma
coupled to the external Lorentz contracted color fields). Common distortion effects due to evolution in finite expanding plasma geometries
will be reported elsewhere.

We use natural units below and Lorentz indices are denoted with Greek
letters $\mu,\nu=0,\ldots,3$ while internal $SU(N_c)$ adjoint color
indices are $a=1,\ldots,N_c^2-1$. Also, the Minkowski metric
$g_{\mu\nu}={\rm diag}(-,+,+,+)$ is employed. In our system of
coordinates, the beam is in the $z$ direction, the associated heavy
quark jet moves along the $x$ direction with velocity ${\bald
v}=v\,\hat{x}$. We define $x_1=x-vt$ and $x_p$ is the transverse
cylindrical radial coordinate perpendicular to the jet axis.

\begin{center}
\begin{figure}[hb]
\centering
\hspace*{-0.4cm}
\epsfig{file=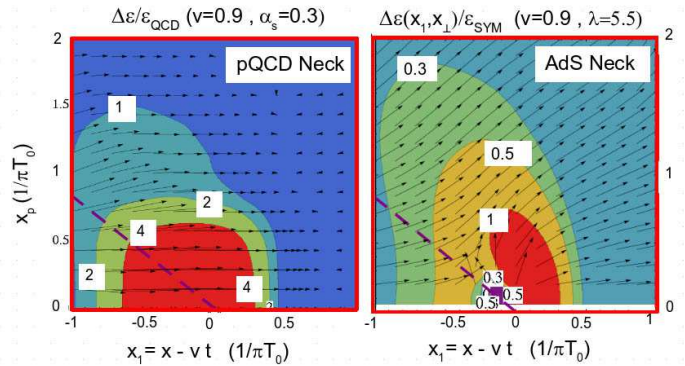,width=3.6in,clip=}
\caption{\label{Stress} (Color online) A magnified view of the
near ``Neck'' zone shows the relative local energy
density perturbation $\Delta \varepsilon/\varepsilon_0$ and fluid flow directions induced by
a heavy supersonic quark jet moving with $v=0.9$. As in Fig.1, the
pQCD contours were computed using 3+1D hydrodynamics
{\protect\cite{Rischke:1995ir}} sourced by \cite{Neufeld:2008hs} (left
panel). The AdS/CFT Neck zone \cite{NGTnonmach} (right panel)
uses numerical tables from \cite{gubsermach} . The purple dashed line indicates the ideal far zone point
source shock angle.  The heavy quark is at the origin of these comoving
coordinates. The arrows indicate both direction and relative magnitude of the fluid
flow velocity. The numbers in the plot label the contours of constant $\Delta
\varepsilon/\varepsilon_0$. Note that $\Delta
\varepsilon/\varepsilon_0$ is larger in pQCD but that the transverse flow
generated near the quark is much stronger in the AdS/CFT model. }
\label{pQCDzones}
\end{figure}
\end{center}

\section{Stress Zones}

Energetic back-to-back jets produced in the early stages of a heavy ion collision transverse to the beam axis  traverse the sQGP and
 deposit energy and momentum
along their path in a way that depends on the non-equilibrium details of the
physics of the field-plasma
coupling. In the case when one of the jets is produced near the surface (the trigger jet), the other supersonic away-side jet  moves
through the plasma and generates in the far zone
a Mach like conical perturbation in the plasma stress as seen in Fig.1.
The resulting conical correlation (with respect to the away-side jet axis)
 is naturally expected to lead to an enhancement of associated
hadrons at the characteristic Mach angle
\cite{Stoecker:2004qu,shuryakcone}.
Below some transverse momentum saturation scale \cite{Mueller:2008zt},
jet physics depends on the properties of the medium
and one can test different models of jet-medium coupling dynamics
by studying the detailed angular and rapidity correlations.
In this work we demonstrate a striking difference
between strongly coupled AdS/CFT and  moderate coupling,
multiple collision pQCD transport models for that coupling, and study their
experimentally neasurable consequences.

We ignore here the  near-side associated correlations and focus on
away-side jet-hadron azimuthal correlations.
The energy-momentum stress induced by the away-side heavy quark jet in both pQCD and AdS/CFT can be conveniently decomposed into four separate
contributions as in \cite{NGTnonmach}
\begin{equation}
T^{\mu\nu}(X)=T^{\mu\nu}_{bg}+\delta T^{\mu\nu}_{Mach}(X)
+ \delta T^{\mu\nu}_{Neck}(X)+\delta T^{\mu\nu}_{Coul}(X)\,.
\label{energymomentum}
\end{equation}
The static isotropic background stress tensor is assumed to be $T^{\mu\nu}_{bg}={\rm diag}(\varepsilon,p,p,p)$,
where $\varepsilon=K T_0^4$ is the corresponding background energy density of a gas of massless $SU(3)$ gluons $K_{QCD}=8\pi^2/15$ whereas for $SU(N_c)$ SYM $K_{SYM}=3\pi^2 (N_c^2-1)/8$. In both cases, $\varepsilon=3\,p\,$ and the background temperature is $T_0$.

The Coulomb contribution to the energy-momentum tensor $\delta
T^{\mu\nu}_{Coul}(X)$ arises from the near zone Lorentz contracted
Coulomb field that remains attached to the heavy quark since we
consider only moderate but supersonic velocities $c_s\le v\le 0.9$. We
can therefore neglect radiative energy loss that dominate in the
ultrarelativistic case.  The bare comoving Coulomb self-field stress has
the singular form $\delta T^{\mu\nu}_{Coul}\propto 1/R^4$ in the quark
rest frame. In both pQCD and AdS/CFT cases we subtract this vacuum
self-field stress as in \cite{gubsermach}. In other words, the zero
temperature contribution to the in-medium stress tensor is always
subtracted. While in AdS/CFT the form of the Coulomb tensor is known
exactly \cite{Friess:2006fk}, in pQCD this contribution can only be
calculated perturbatively. The leading order expression for the
chromo-fields produced by the source in pQCD, in the limit where the
dielectric functions are set to unity, displays the same
Lienard-Wiechert behavior as in AdS/CFT.

The far zone ``Mach'' part of the stress can be expressed in terms of the  local temperature $T(X)$ and fluid flow velocity fields
$U^\alpha(X)$ through the first-order Navier-Stokes stress form
\begin{eqnarray}
\delta T_{Mach}^{\mu\nu}(X)  &=&  \left[sT\left(U^\mu U^\nu+\frac{3}{4}g^{\mu\nu} -\frac{\eta}{sT} \partial^{\{ \mu}U^{\nu\}}
\right)\right. \nonumber \\ &-& \left.T^{\mu\nu}_{bg}\right] \,\theta(1-3K_N)
\label{hydrotensor}
\end{eqnarray}
where $s(X)$ is the local entropy density and $\partial^{\{ \mu}U^{\nu\}} $ is the symmetrized traceless shear flow velocity gradient. In AdS/CFT, $\eta/s=1/4\pi$ \cite{Policastro:2001yc}. The theta function in Eq.\ (\ref{hydrotensor}) defines the ``far zone'' that includes the Mach and Diffusion linearized hydrodynamic sound waves. In the far zone
equilibration
rate is over three times the local stress gradient scale, and first-order Navier-Stokes dissipative hydrodynamics provides an adequate description of the evolution in that zone.
In AdS/CFT, the hydrodynamic description was shown to be valid down to distances of roughly $3/\left(\pi T\right)$
\cite{Noronha:2007xe} (see also
\cite{Chesler:2007sv} for an analysis of the far zone).
The near Neck zone within a thermal Compton wavelength of the heavy quark
is a generally nonequilibrium region  strongly influenced by
the coupling of the heavy quark's bare classical Coulomb field to the polarizable plasma.

\section{The pQCD source term}

We use the nonlocal pQCD chromo-viscous stress source, $S^\nu(x)$,
 derived in Ref.\ \cite{Neufeld:2008hs}, to drive the perfect fluid
response of a pQCD fluid assuming the ideal $\eta=0$
in order to maximize any pQCD transport induced
azimuthal conical signature. Finite viscosity of course washes out some of
the induced correlations as shown in \cite{Neufeld:2008dx}. However, our main
finding below is that even in this perfect $\eta=0$ hydrodynamic
limit the induced correlations sourced in pQCD are too weak
to generate conical correlations after freeze-out.

 Thus, we consider only the $\eta/s=0$ limit of the full anomalous chromo-viscous equations derived in
\cite{Asakawa:2006jn}. However, we retain the anomalous diffusion stress Neufeld source (Eqs.\ (23-24) of \cite{Neufeld:2008hs}). We can rewrite  Eqs.\ (6.2 - 6.11) of \cite{Asakawa:2006jn} in the more familiar covariant Joule heating form
\begin{eqnarray}
\partial_\mu T^{\mu\nu}=\mathcal{S}^\nu= F^{\nu\alpha\,a}J_{\alpha}^a&=&
(F^{\nu\alpha\,a}\sigma_{\alpha\beta\gamma}*F^{\beta\gamma\,a})
\end{eqnarray}
where $F^{\mu\nu\,a}(X)$ is the external Yang-Mills field tensor and $J^a(X)=\int d^4 K/(2\pi)^4  \exp(-iK\cdot X) J^a(K)$ is the color current that is related via
Ohm's
law to $F^{\mu\nu\,a}(K)$ through the (diagonal in color) conductivity rank three tensor $\sigma_{\mu\alpha\beta}(K)$. The $*$ denotes a convolution over the
nonlocal
non-static conductive dynamical response of the polarizable plasma.
The covariant generalization of Neufeld's source is most easily understood
 through its Fourier
decomposition, $J_{\nu}^a(K)=\sigma_{\nu\mu\alpha}(K)
F^{\mu\alpha\,a}(K)$, with the color conductivity expressed as
\cite{Selikhov:1993ns}
\begin{equation}
\sigma_{\mu\alpha\beta}(K)=i g^2\int d^4 P\frac{P_\mu P_\alpha\,\partial^P_\beta}{
P\cdot K+i \,P\cdot U/\tau^*}f_0(P)
\label{conduct}
\end{equation}
where $f_0(P)=2\left(N_c^2-1\right) \,G(P)$ is the effective
plasma equilibrium distribution with $G(P)=(2\pi^3)^{-1}\theta(P_0)\delta(P^2)/(e^{P_0/T}-1)$. Here, $U^\mu$ is the 4-velocity of the plasma as in Eq.\
(\ref{hydrotensor}).
For an isotropic plasma $U^{\beta}\sigma_{\mu\alpha\beta}(K)=-\sigma_{\mu\alpha}(K)$. In the long wavelength limit,
$U_{\beta}\sigma^{\mu\alpha\beta}(K\rightarrow 0) =-\tau^* m_D^2 \,g^{\mu\alpha} /3$, where $m_D^2=g^2 T^2$ is the Debye screening mass for a
noninteracting plasma of
massless $SU(3)$ gluons in thermal equilibrium.

The relaxation  or decoherence time $\tau^*$ in (\ref{conduct})
has the general form noted in \cite{Asakawa:2006jn}
${\tau^*}= (1/\tau_p + 1/{\tau_c} + {1}/\tau_{an})^{-1}$
where $\tau_p\propto (\alpha_s^2 T\ln (1/\alpha_s))^{-1}$ being the collisional momentum relaxation time \cite{Danielewicz:1984ww,Heinz:1985qe},
$\tau_c=(\alpha_s N_c T\ln (1/g))^{-1}$ being the color diffusion time defined in
\cite{Selikhov:1993ns}, and $\tau_{an}\propto (m_D(\eta|\bald{\nabla}\cdot \bald{U}|/Ts)^{1/2})^{-1}$ being the anomalous strong electric and
magnetic
field relaxation time derived in Eq.\ (6.42) of \cite{Asakawa:2006jn}. We
note that one can express
\begin{equation}
\tau_{an}\propto\frac{1}{gT} \frac{1}{\surd K_N(X)}
\label{anom}
\end{equation}
in terms of the local Knudsen number $K_N=\Gamma_s/L$ used in Eq.\ (\ref{hydrotensor}). Here $L$ is the
characteristic stress gradient scale. However, because $\eta\propto\tau^* sT$, Eq.\ (\ref{anom}) is really an implicit equation for $\tau_{an}$. Combining these relations and
taking into account the uncertainty principle constraint \cite{Danielewicz:1984ww} that bounds $\tau^*\stackrel{>}{\sim} 1/\left(3T\right)$ for an ultrarelativistic
(conformal) plasma, we have
\begin{equation}
 \frac{1}{\tau^*} \propto T \left(a_1\; g^4 \ln g^{-1}
+a_2\; g^2\ln g^{-1} + a_{3}\; g \surd
K_N \right) \stackrel{<}{\sim} 3T
\label{rate}
\end{equation}
where $a_1\,,a_2\,,a_3$ are numerical factors. In the near zone (see Fig.2)
close to the quark, $K_N$ gets large, $\tau_{an}$ can become the dominant
contribution to $\tau^*$  in the presence of strong classical field gradients.

We would like to emphasize an important subtle point in the application of
Eq.\ (\ref{rate}) to our heavy quark jet problem.
In order to neglect viscous dissipation in the pQCD response,
the relaxation rate must be very large compared to the characteristic
gradient scale. Hence, in the far zone at least
the imaginary part of the conductivity denominator
in Eq.(\ref{conduct}) must be large and dominant.
However, in the Neck region the field gradients become
 very large and the relevant
wave numbers of the hydro response  $K\gg 3T$ exceed the uncertainty
limited equilibration rate. Because we only need to consider
the conductivity in the asymptotic
large $K$ limit in the near zone, it becomes possible
to neglect the $\sim i 3T$ maximal relaxation rate in the energy denominator
and to formally set  $1/\tau^* \rightarrow 0^+$ - {\em as if}
 the coupling were parametrically small (as assumed in Eqs.\ (53-56) of \cite{Neufeld:2008hs}).
 Only in this high frequency, high wave number limit,
relevant for the Neck zone physics, is
the color conductivity computable as in \cite{Neufeld:2008hs}.
The neglect of dissipation in the Neck zone
maximizes the acceleration of the plasma partons, which can subsequently generate transverse collective plasma flow relative to the jet axis. What we must check numerically is whether
this maximum transfer of field energy-momentum from the field to the plasma
is sufficiently anisotropic to generate a conical correlation
of the associated hadron fragments.

\section{Freeze-out Procedures}

As noted previously, we consider here only the
idealized static medium here to maximize plasma response signals.
Distortion effects due transverse expansion and non-conformal equation of states with a phase transition will be presented elsewhere.
These effects, while important for phenomenological comparisons
to heavy ion data, however obscure the fundamental differences between
weak and strong coupled dynamics that is our focus here.

Given the large theoretical systematic uncertainty inherent
in any phenomenological model of nonperturbative hadronization,
we consider here two simple limits for modeling
the fluid decoupling and freeze-out.
In one often used limit, we freeze-out computational fluid
cell via the CF prescription on an isochronous hypersurface.
This scheme takes into account maximal thermal broadening effects.
In the opposite limit, we freeze-out assuming isochronous sudden breakup or shattering
of fluid cells conserving only energy and momentum and avoiding hadronization
altogether as described in more detail below. The difference between the two schemes provides a measure of the systematic theoretical uncertainty associated with the unsolved problem of hadronization.

In the CF method, the conversion of the fluid into free particles is achieved instantaneously at a critical surface $d \Sigma_\mu$ \cite{Cooper:1974mv}. If we assume such a
freeze-out scheme
\cite{shuryakcone,NGTnonmach,Betz:2008js,heinzcone}, the particle distributions and correlations can be obtained from the flow velocity field $U^{\mu}(X)$
and temperature
profile $T(X)$. 
For associated (massless) particles with
$P^{\mu}=(p_{T},p_{T}\cos (\pi-\phi),p_{T}\sin (\pi-\phi),0)$ the momentum distribution at mid rapidity $y=0$ is
\begin{equation}
\frac{dN}{p_Tdp_Tdy d\phi}\Big
|_{y=0}=\int_{\Sigma}d\Sigma_{\mu}P^{\mu}\left[f_0(U^{\mu},P^{\mu},T)-f_{eq}\right]
\label{cooperfrye}
\end{equation}
where $p_T$ is the transverse momentum, $\Sigma (X)$ is the freeze-out hypersurface, and $f_0=\exp(-U^{\mu}P_{\mu}/T(X))$ is a local Boltzmann
equilibrium
distribution. No viscous corrections to Eq.\ (\ref{cooperfrye}) are included
since we are working here in the perfect fluid limit with $\eta=0$.
We subtract  the isotropic background yield via
$f_{eq}\equiv f|_{U^{\mu}=0,T=T_0}$. Moreover, we follow \cite{shuryakcone,NGTnonmach,Betz:2008js} and perform an isochronous freeze-out where
$d\Sigma^\mu= d^3 \,{\bald x} \left(1,0,0,0\right)$.

We remark that the absence of well-defined quasi-particle states in AdS/CFT plasmas at large t'Hooft coupling indicates that CF can only, at best, give a qualitative idea of the observable hadron level angular correlations \cite{NGTnonmach}. Moreover, even in the pQCD quasiparticle limit, CF freeze-out remains a strong model assumption. In the pQCD case, in the associated $p_T$ range of interest a coalescence/recombination hadronization scenario \cite{Fries:2003vb,Fries:2004hd} may be more appropriate. However, we expect similar CF thermal broadening effects if coalescence hadronization is assumed and full three momentum conservation is taken into account.

As an alternate freeze-out scheme we consider a calorimetric-like
observable 
given by the momentum density weighted 
polar angle distribution relative to the jet axis:
\begin{eqnarray}
\hspace{-2in}\frac{d S}{d\cos\theta} &=& \sum_{cells} |\vec{\mathcal{P}}_c|
\delta\left(\cos\theta- \cos\theta_c\right)\nonumber \\
&=& \int d^3 {\bf x}\,\, |\bald{M}(X)|
\delta\left(\cos\theta- \frac{M_x(X)}{|{\bf M}(X)|}\right)\Big|_{t_f}
\label{bulkeq}
\end{eqnarray}
 This quantity differs from CF mainly by the neglect of the thermal
 smearing at the freezeout time, and thus it maximally amplifies the angular
 anisotropy of the associated hadrons. 
The very strong assumption in this decoupling
scheme is that hadrons from each frozen-out cell
emerge parallel to the cell total momentum
$\mathcal{P}^i_c=d^3{\bf x}\,\, T^{0i}({\bf x},t_f)$.
Here $\theta=\pi-\theta_{trigger}$ is the polar angle
with respect to the away-side heavy quark jet.
Many other similar purely hydrodynamic measures of bulk
flow are possible \cite{Stoecker:2004qu}, e.g. entropy instead of
momentum density weighted. However, we found no qualitative
differences when the weight function is changed. We used a narrow Gaussian
approximation to the Dirac delta in Eq.\ (\ref{bulkeq})  with 
a $\Delta\cos\theta=0.05$ width and checked that the results 
did  not change significantly if the width was
varied by $50\%$.

Our goal in this letter is not to decide which hadronization scheme is
preferred but to apply these two commonly used measures of
hydrodynamic to help quantify the observable differences between two
strongly different approaches to jet-plasma interactions (pQCD x
AdS/CFT). The CF freeze-out employed here and in
Refs.\ \cite{shuryakcone,NGTnonmach,Betz:2008js,heinzcone} is
especially questionable in 
the non-equilibrium Neck region but provides a rough estimate
of intrinsic thermal smearing about the local hydrodynamic flow.
  In Fig.\ 2 we show the relative local energy
disturbance and flow profile in the Neck region created by a v=0.9 jet
in both pQCD and AdS/CFT. Note that the relative transverse flow in
the Neck zone in AdS/CFT is significantly larger than in pQCD and as
we show below this is reflected in the final angular correlations from that
region in both hadronization schemes.

\section{Freeze-out results in pQCD}

The initial away-side heavy quark jet is assumed to start
 $t=0$ at $x_1 =-4.5$ fm and the freeze-out is done
when the heavy quark reaches the origin of the coordinates at time
$t_f=4.5/v$ fm. This provides a rough description
of the case in which a
uniformly moving heavy quark punches through the
medium after passing through $4.5$ fm of plasma.

The numerical output of the SHASTA code is the temperature and fluid flow velocity fields $T(X)$ and ${\bf U}(X)$. The hydrodynamic equations were solved in the presence of the source term $\mathcal{S}^{\mu}(X)$ computed analytically by Neufeld in \cite{Neufeld:2008hs} in the limit where the dielectric functions that describe the medium's response to the color fields created by the heavy
quark were set to unity.
The effects from medium screening on $\mathcal{S}^{\mu}$ were studied in detail in Ref.\ \cite{Neufeld:2008hs}. In our
numerical calculations we used $x_{p\,max}=1/m_D$ as an infrared cutoff while the minimum lattice spacing naturally provided an ultraviolet cutoff.
The background temperature was set to $T_0=0.2$ GeV. We assumed $\alpha_s=g^2/\left(4\pi\right)=1/\pi$ in our calculations involving the pQCD source.

\begin{center}
\begin{figure}[t!]
\centering
\epsfig{file=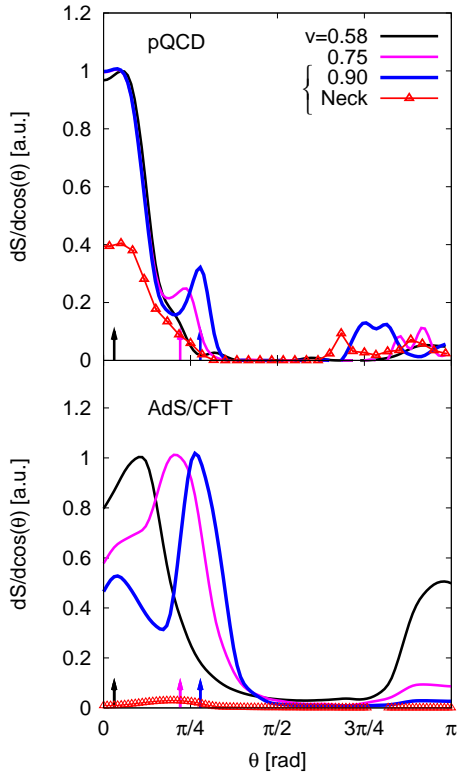,width=2.6in,clip=}
\caption{\label{bulkmomentum}
(Color online) The (normalized) momentum weighted bulk flow angular distribution as a function of polar angle with respect to the away-side jet is shown for $v=0.58$ (black), $v=0.75$ (magenta), and $v=0.90$ (blue) comparing pQCD anomalous chromo-hydrodynamics to the AdS/CFT string drag
\cite{gubsermach,Herzog:2006gh} model analyzed in Ref.\ {\protect\cite{NGTnonmach}}. The red line with triangles represents the Neck contribution for a jet with $v=0.9$ and the arrows indicate the location of the ideal Mach-cone angle given by $\cos\theta_M=c_s/v$, where
$c_s=1/\sqrt{3}$. }
\end{figure}
\end{center}

The results for the bulk flow according to Eq.\ (\ref{bulkeq}) in pQCD
are shown in the upper panel in Fig.\ \ref{bulkmomentum}. The curves
are normalized in such a way that the largest contributions are set to
unity. Note that the pQCD bulk energy flow distribution has a large
forward moving component in the direction of the jet for all the
velocities studied here. In the far zone, this forward moving energy
flow corresponds to the diffusion wake studied in
\cite{shuryakcone}. The red curve with triangles in the upper panel in
Fig.\ \ref{bulkmomentum} corresponds to the yield solely from the Neck
region for $v=0.9$. The very small dip at small $\theta=0$ is mostly
due to the weak Neck zone pQCD but the most of the momentum flow from
both Neck and Diffusion wakes is directed along the jet axis. The
relatively small transverse energy flow in the Neck region is evident
on the left panel of see Fig. 2 in contrast to the much larger
transverse flow predicted via AdS in that near zone.  The Mach cone
emphasized in Ref.\ \cite{Neufeld:2008fi, Neufeld:2008dx} is
also clearly seen but its amplitude relative to the mostly forward
diffusion plus Neck contribution is much smaller than in the AdS/CFT
case. The weak Mach peak roughly follows Mach's law as $v$
approaches $c_s$.


In Fig. 4, our CF freeze-out
results for the associated away-side azimuthal distribution
for $v=0.58,0.75,0.9$ for  mid-rapidity and $p_T=5 \pi \,T_0 \sim 3.14$ GeV light hadrons are shown. The pQCD case , computed using the output of the SHASTA code into Eq.\ (\ref{cooperfrye}), are shown in the upper panel. We show the angular function
\begin{equation}
\ CF(\phi)=\frac{1}{N_{max}}\left(\frac{dN(\phi)}{p_Tdp_Tdy d\phi}\right)\Big|_{y=0}
\label{CFfunction}
\end{equation}
where $N_{max}$ is a constant used to normalize the plots (note that
this function is not positive-definite). The pQCD angular distribution
shows only a sharp peak at $\phi=\pi$ for all velocities. The red
curve with triangles denotes the contribution from the pQCD Neck
region for $v=0.9$. Note that the different peaks found in the bulk
flow analysis of the pQCD data shown in the upper panel in
Fig. \ref{bulkmomentum} do not survive CF freeze-out. We checked that
no other structures appear if we either double $p_T$ to 5 GeV or
increase $\alpha_s$ to 0.5. We conclude that the strong forward moving diffusion wake
as well as the mostly forward bow shock Neck zone dominate
the away-side peak and that the thermal broadened Mach correlations are too weak in pQCD to contribute to the final angular correlations.

\section{Freeze-out Results in AdS/CFT}

We used the same setup employed in \cite{NGTnonmach} to perform the CF
freeze-out of the $\mathcal{N}=4$ SYM AdS/CFT data computed by Gubser,
Pufu, and Yarom in Ref.\ \cite{gubsermach}. They calculated the
energy-momentum disturbances caused by the heavy-quark, which in this
steady-state solution was created at $t\to -\infty$ and has been
moving through the infinitely extended $\mathcal{N}=4$ SYM static
background plasma since then. The freeze-out is computed when the
heavy quark reached the origin of the coordinates. The mass of the
heavy quark $M$ in the AdS/CFT calculations is such that $M/T_0\gg
\sqrt{\lambda}$, which allows us to neglect the fluctuations of the
string \cite{Herzog:2006gh,CasalderreySolana:2007qw}. At $N_c=3$ the
simplifications due to the supergravity approximation are not strictly
valid but it is of interest to extrapolate the numerical solutions to
study its phenomenological applications.  We chose a plasma volume to
be the forward light-cone that begins at $x_1=-4.5$ fm and has a
transverse size of $x_p< 4.5$ fm at $T_0=0.2$ GeV (our background
subtracted results do not change when larger volumes were used).  Note
that we assumed the same background temperature for both pQCD and
AdS/CFT.  The mapping between the physical quantities in
$\mathcal{N}=4$ SYM and QCD is a highly non trivial open problem (see,
for instance, the discussion in Ref.\ \cite{gubserlambda}). We
therefore again use CF and bulk momentum flow as two extreme limits to
gauge possible systematic uncertainties.

The (normalized) bulk
momentum flow associated with the AdS/CFT data, computed using Eq.\
(\ref{bulkeq}), is shown in the lower panel of Fig.\
\ref{bulkmomentum}. The bottom panel in Fig.\ \ref{bulkmomentum}
simply shows that in AdS/CFT there are more cells pointing in a
direction near the Mach cone angle than in the forward direction
(diffusion wake) when $v=0.9$ and $v=0.75$ unlike in the pQCD case shown in the upper panel. However, when $v=0.58$ the
finite angle from the Mach cone is overwhelmed by the strong bow shock
formed in front of the quark, which itself leads to small conical dip
not at the ideal Mach angle (black arrow).
 The red line with triangles in the
bottom panel of Fig.\ \ref{bulkmomentum} shows that the relative
magnitude of the contribution from the Neck region to the final bulk
flow result in AdS/CFT is much smaller than in pQCD. However, note
that small amplitude peak in the AdS/CFT Neck curve is located at a
much larger angle than the corresponding peak in the pQCD Neck, as one
would expect from the transverse flow shown Fig.\ 2.
Moreover, one can see that a peak in the
direction of the trigger particle can be found for all the velocities
studied here. This peak represents the backward flow that is always
present vortex-like structures created by the jet as
discussed in detail in Ref.\ \cite{Betz:2007kg}. Possible phenomenological
consequences due to these vortex structures on  the polarization of
hyperons in heavy ion collisions were discussed in Ref.\
\cite{Betz:2007kg}.

\begin{figure}[t]
\centerline
\centering
\epsfig{file=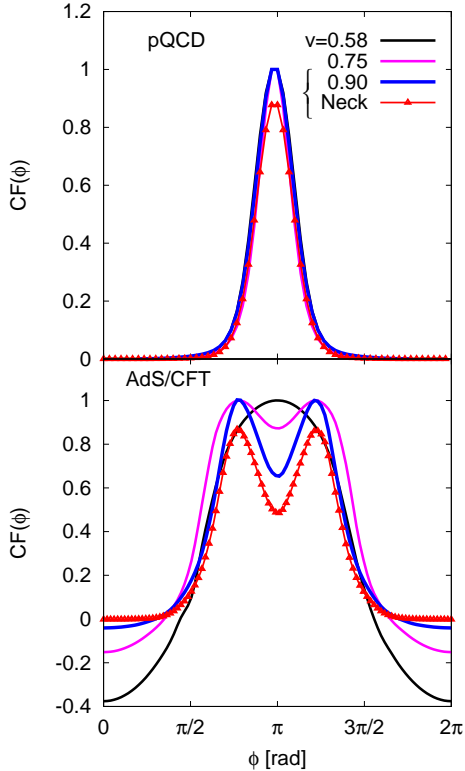,width=2.6in,clip=}
\caption{\label{CFplot}
(Color online) Normalized (and background subtracted) azimuthal away-side jet associated correlation after Cooper-Frye freeze-out $CF(\phi)$ (see Eq.\ \ref{CFfunction}) for pQCD (top) and AdS/CFT from {\protect\cite{NGTnonmach}} (bottom). Here $CF(\phi)$ is evaluated at $p_T=5 \pi \,T_0 \sim 3.14$ GeV and $y=0$. The black line is for $v=0.58$, the magenta line for $v=0.75$, and for the blue line $v=0.9$. The red line with triangles represents the Neck contribution for a jet with $v=0.9$.}
\end{figure}

Our results for the CF freeze-out of the AdS/CFT solution for
$v=0.58,0.75,0.9$ at mid-rapidity and $p_T=5 \pi T_0$ GeV in the lower
panel in Fig.\ \ref{CFplot}. A double peak structure can be seen for
$v=0.9$ and $v=0.75$. Note, however, that the peaks in the AdS/CFT
correlation functions do not obey Mach's law. This is because these
correlations come from the Neck region where there is a strong
transversal non-Mach flow \cite{NGTnonmach}. This is explicitly shown
by the red curve with triangles that represents the Neck contribution
for a jet with $v=0.9$ as in Fig.\ \ref{bulkmomentum}. For $v=0.58$,
the resulting flow is not strong enough to lead to non-trivial angular
correlations. A detailed study of the $p_T$ dependence of the
away-side correlations associated with AdS/CFT heavy quark jets will
be presented in a future work. In addition, the negative yield present
in the CF curves for $v=0.58$ and $v=0.75$ is due to the presence of
the vortices discussed above.

In general \cite{NGTnonmach}, the weak sound waves produced by a jet
do not lead to a cone-like signal independently of the detailed flow
and interference patterns because thermal smearing washes out the
signal. Formally, if linearized hydrodynamics applies and in the low
momentum limit (${\bald U}\cdot {\bald p}<<T$), the associated hadron
away-side distribution is only a very broad peak about
$\phi=\pi$ regardless of the detailed combination of Mach wakes,
diffusion wakes, or vortex circulation
\cite{shuryakcone,NGTnonmach,Betz:2007kg}. This result involving CF
freeze-out can only be circumvented either in regions with high flow
velocities and large gradients as in the Neck zone \cite{NGTnonmach},
or by increasing $p_T$ to unrealistic high values \cite{shuryakcone,Betz:2008js}.

One of the main differences between the two freeze-out procedures we
employed (in both AdS/CFT and pQCD) concerns the relative magnitude of
the contribution from the Neck region to the final angular
correlations: the Neck region is much more important in CF than in the
bulk flow measure computed via Eq.\ (\ref{bulkeq}). This is due to the
exponential factor in CF, which largely amplifies the contribution
from the small region close to the jet where the disturbances caused
by the heavy quark become relevant.

\section{Conclusions}

In this paper, we showed that the angular correlations obtained after
an isochronous Cooper-Frye freeze-out of the wake induced by
punch-through heavy quark jets (in a static medium) in the Neufeld
{\it et al.} pQCD model of anomalous chromo-viscous hydrodynamics do
not display a conical structure. This should be compared to the
conical-like structures seen after CF freeze-out of the
strongly-coupled AdS/CFT string drag model. We expect similar thermal
broadening effects to occur if alternative coalescence/recombination
hadronization models \cite{Fries:2003vb,Fries:2004hd} are used. The
isochronous hypersurface we used is needed in order to compare AdS/CFT
to pQCD since AdS/CFT heavy quark solutions have only been computed so
far in a static medium. For realistic simulations that can be compared
to data, effects from the medium's longitudinal, transverse, and
elliptic flow must be taken into account as discussed in detail in
\cite{heinzcone}.

Unlike AdS/CFT, the conical flow from the associated nonequilibrium
Neck zone in pQCD (see the red region in the left panel of Fig.\ 2 and
the red curve in Fig.\ 3) is too weak to survive CF freeze-out. In
both cases, the actual Mach wakes do not appear after standard CF
freeze-out.  Mach-like peaks are only observable
in the sudden shattering  freeze-out scenario described in Eq.\
(\ref{bulkeq}) in both pQCD and AdS/CFT, in which
thermal broadening is entirely neglected.

 The Neck region (in both pQCD and AdS/CFT) gives
the largest contribution to the total yield in CF freeze-out while its
contribution in the other extreme case involving the bulk flow
hadronization is not as relevant. This indicates that the magnitude of
the Neck's contribution to the final angular correlations is still
strongly model dependent. Nevertheless, our results suggest
that conical but non-Mach law correlations are
much more likely to appear in AdS/CFT than in pQCD.

We propose that the measurement of the jet velocity dependence of the
associated away-side correlations with identified heavy quark triggers
at RHIC and LHC will provide important constraints on possible pQCD
versus AdS/CFT dynamical non-Abelian field - plasma (chromo-viscous)
coupling models.


\section*{Acknowledgments}

We thank S.\ Gubser, A.\ Yarom, and S.\ Pufu for providing numerical AdS/CFT stress tables, D.\ Rischke for permission to use his SHASTA code. The authors thank W.\ Zajc, B.\ M\"uller, M.\ Strickland, J.\ Ulery, R.\ Neufeld, W.\ Horowitz, H.\ St\"ocker, D.\ Rischke, C.\ Greiner, J.\ Reinhardt, and I.\ Mishustin for valuable comments and suggestions. J.N. and M.G. acknowledge support from DOE under Grant No. DE-FG02-93ER40764. M.G. acknowledges sabbatical support from DFG, ITP, and FIAS at Goethe University. G.T. thanks the Alexander Von Humboldt foundation and Goethe University for support.



\begin{thebibliography}{99}

\bibitem{Adler:2005ee}
  S.~S.~Adler {\it et al.}  [PHENIX Collaboration],
  Phys.\ Rev.\ Lett.\  {\bf 97}, 052301 (2006);
  J.~Adams {\it et al.}  [STAR Collaboration],
  Phys.\ Rev.\ Lett.\  {\bf 95}, 152301 (2005);
  J.~G.~Ulery  [STAR Collaboration],
  Nucl.\ Phys.\  A {\bf 774}, 581 (2006);
  A.~Adare {\it et al.}  [PHENIX Collaboration],
  Phys.\ Rev.\  C {\bf 78}, 014901 (2008); 
   B.~I.~Abelev {\it et al.}  [STAR Collaboration],
  arXiv:0805.0622 [nucl-ex].


\bibitem{Stoecker:2004qu}
  H.~Stoecker,
  Nucl.\ Phys.\  A {\bf 750}, 121 (2005).

\bibitem{shuryakcone}
J.~Casalderrey-Solana, E.~V.~Shuryak and D.~Teaney,
  Nucl.\ Phys.\  A {\bf 774}, 577 (2006);
  arXiv:hep-ph/0602183.


\bibitem{Gyulassy:2004zy}
  M.~Gyulassy and L.~McLerran,
  Nucl.\ Phys.\  A {\bf 750}, 30 (2005); E.~V.~Shuryak,
  Nucl.\ Phys.\  A {\bf 750}, 64 (2005).

\bibitem{NGTnonmach}
J.~Noronha, M.~Gyulassy, and G.~Torrieri, arXiv:0807.1038 [hep-ph];
  M.~Gyulassy, J.~Noronha, and G.~Torrieri, arXiv:0807.2235 [hep-ph];
J.~Noronha and M.~Gyulassy,
  arXiv:0806.4374 [hep-ph];



\bibitem{Herzog:2006gh}
  C.~P.~Herzog, A.~Karch, P.~Kovtun, C.~Kozcaz and L.~G.~Yaffe,
  JHEP {\bf 0607}, 013 (2006);
  S.~S.~Gubser,
  PRD {\bf 74}, 126005 (2006).

\bibitem{gubsermach}
  S.~S.~Gubser, S.~S.~Pufu and A.~Yarom,
  Phys.\ Rev.\ Lett.\  {\bf 100}, 012301 (2008);
arXiv:0711.1415 [hep-th];
 JHEP {\bf 0709}, 108 (2007).

\bibitem{maldacena}
J.~M.~Maldacena,
  Adv.\ Theor.\ Math.\ Phys.\  {\bf 2}, 231 (1998)
  [Int.\ J.\ Theor.\ Phys.\  {\bf 38}, 1113 (1999)]; E.~Witten,
  Adv.\ Theor.\ Math.\ Phys.\  {\bf 2}, 253 (1998);
  {\bf 2}, 505 (1998);
  O.~Aharony, S.~S.~Gubser, J.~M.~Maldacena, H.~Ooguri and Y.~Oz,
  Phys.\ Rept.\  {\bf 323}, 183 (2000).

\bibitem{Friess:2006fk}
  J.~J.~Friess, S.~S.~Gubser, G.~Michalogiorgakis and S.~S.~Pufu,
  Phys.\ Rev.\  D {\bf 75}, 106003 (2007).

\bibitem{Horowitz:2007su}
  W.~A.~Horowitz and M.~Gyulassy,
  arXiv:0706.2336 [nucl-th].

\bibitem{Chesler:2007an}
  P.~M.~Chesler and L.~G.~Yaffe,
  Phys.\ Rev.\ Lett.\  {\bf 99}, 152001 (2007).

\bibitem{Policastro:2001yc}
  G.~Policastro, D.~T.~Son and A.~O.~Starinets,
  Phys.\ Rev.\ Lett.\  {\bf 87}, 081601 (2001); 
  P.~Kovtun, D.~T.~Son and A.~O.~Starinets,
  Phys.\ Rev.\ Lett.\  {\bf 94}, 111601 (2005).

\bibitem{Danielewicz:1984ww}
  P.~Danielewicz and M.~Gyulassy,
  Phys.\ Rev.\  D {\bf 31}, 53 (1985).
%

\bibitem{Gubser:2007ni}
  S.~S.~Gubser and A.~Yarom,
  Phys.\ Rev.\  D {\bf 77}, 066007 (2008);
  arXiv:0803.0081 [hep-th].

\bibitem{Cooper:1974mv}
  F.~Cooper and G.~Frye,
  Phys.\ Rev.\  D {\bf 10}, 186 (1974).

\bibitem{Noronha:2007xe}
  J.~Noronha, G.~Torrieri and M.~Gyulassy,
  arXiv:0712.1053 [hep-ph], PRC in press; J.~Noronha, M.~Gyulassy and G.~Torrieri,
  arXiv:0806.4665 [hep-ph], JPG in press.

\bibitem{Yarom:2007ni}
  A.~Yarom,
  Phys.\ Rev.\  D {\bf 75}, 105023 (2007).

\bibitem{Gubser:2007nd}
  S.~S.~Gubser and S.~S.~Pufu,
  Nucl.\ Phys.\  B {\bf 790}, 42 (2008).


\bibitem{Neufeld:2008fi}
  R.~B.~Neufeld, B.~Muller and J.~Ruppert,
  Phys.\ Rev.\  C {\bf 78}, 041901(R) (2008)
  [arXiv:0802.2254 [hep-ph]].


\bibitem{Neufeld:2008hs}
  R.~B.~Neufeld,
  Phys.\ Rev.\  D {\bf 78}, 085015 (2008)
  [arXiv:0805.0385 [hep-ph]].


\bibitem{Neufeld:2008dx}
  R.~B.~Neufeld,
  arXiv:0807.2996 [nucl-th].

\bibitem{Asakawa:2006tc}
  M.~Asakawa, S.~A.~Bass and B.~Muller,
  Phys.\ Rev.\ Lett.\  {\bf 96}, 252301 (2006).


\bibitem{Asakawa:2006jn}
  M.~Asakawa, S.~A.~Bass and B.~Muller,
  Prog.\ Theor.\ Phys.\  {\bf 116}, 725 (2007).

\bibitem{Heinz:1985qe}
  U.~W.~Heinz,
  Annals Phys.\  {\bf 168}, 148 (1986);
  H.~T.~Elze and U.~W.~Heinz,
  Phys.\ Rept.\  {\bf 183}, 81 (1989).

\bibitem{Rischke:1995ir}
  D.~H.~Rischke, S.~Bernard and J.~A.~Maruhn,
  Nucl.\ Phys.\  A {\bf 595}, 346 (1995); 
  D.~H.~Rischke, Y.~Pursun and J.~A.~Maruhn,
  Nucl.\ Phys.\  A {\bf 595}, 383 (1995)
  [Erratum-ibid.\  A {\bf 596}, 717 (1996)]; 
  D.~H.~Rischke, Y.~Pursun, J.~A.~Maruhn, H.~Stoecker and W.~Greiner,
  Heavy Ion Phys.\  {\bf 1}, 309 (1995);
  S.~Bernard, J.~A.~Maruhn, W.~Greiner and D.~H.~Rischke,
  Nucl.\ Phys.\  A {\bf 605}, 566 (1996).

\bibitem{Betz:2008js}
  B.~Betz, M.~Gyulassy, D.~H.~Rischke, H.~Stocker and G.~Torrieri,
  arXiv:0804.4408 [hep-ph].

\bibitem{Betz:2007kg}
  B.~Betz, M.~Gyulassy and G.~Torrieri,
  Phys.\ Rev.\  C {\bf 76}, 044901 (2007).

\bibitem{Mueller:2008zt}
  A.~H.~Mueller,
  arXiv:0805.3140 [hep-ph].

\bibitem{Dominguez:2008vd}
  F.~Dominguez, C.~Marquet, A.~H.~Mueller, B.~Wu and B.~W.~Xiao,
  arXiv:0803.3234 [nucl-th].

\bibitem{Chesler:2007sv}
  P.~M.~Chesler and L.~G.~Yaffe,
  arXiv:0712.0050 [hep-th].

\bibitem{Selikhov:1993ns}
  A.~Selikhov and M.~Gyulassy,
  Phys.\ Lett.\  B {\bf 316}, 373 (1993);
  Phys.\ Rev.\  C {\bf 49}, 1726 (1994);
  K.~J.~Eskola and M.~Gyulassy,
  Phys.\ Rev.\  C {\bf 47}, 2329 (1993).



\bibitem{heinzcone}
  A.~K.~Chaudhuri and U.~Heinz,
  Phys.\ Rev.\ Lett.\  {\bf 97}, 062301 (2006).

\bibitem{Fries:2003vb}
  R.~J.~Fries, B.~Muller, C.~Nonaka and S.~A.~Bass,
  Phys.\ Rev.\ Lett.\  {\bf 90}, 202303 (2003); Phys.\ Rev.\  C {\bf 68}, 044902 (2003).

\bibitem{Fries:2004hd}
  R.~J.~Fries, S.~A.~Bass and B.~Muller,
  Phys.\ Rev.\ Lett.\  {\bf 94}, 122301 (2005).

\bibitem{CasalderreySolana:2007qw}
  J.~Casalderrey-Solana and D.~Teaney,
  JHEP {\bf 0704}, 039 (2007).

\bibitem{gubserlambda}
S.~S.~Gubser, Phys.\ Rev.\  D {\bf 76}, 126003 (2007).

%
%
%
%







\end{thebibliography}
\end{document}